\documentclass[11pt,american,twoside,a4paper]{article}
\usepackage[T1]{fontenc}
\usepackage[utf8]{inputenc}
\usepackage{latexsym}
\usepackage{amsmath}
\usepackage{bbm}
\usepackage{amssymb}
\usepackage{babel}
\usepackage{amsfonts}
\usepackage{fourier}
\usepackage{theorem}
\usepackage{epsfig}
\usepackage{enumerate}
\usepackage{color,xcolor}
\usepackage{xspace,needspace,xparse,easyReview}
\usepackage[active]{srcltx}
\usepackage[colorlinks=true]{hyperref}
\usepackage{fancyhdr}
\usepackage{xargs}
\usepackage[toc]{appendix}

\newcommandx{\ca}[2][1=]{\todo[inline,author={ca},
	linecolor=green,backgroundcolor=green!15,bordercolor=green,#1]{#2}}
\newcommandx{\canot}[2][1=]{\todo[author={ca},
	linecolor=green,backgroundcolor=green!15,bordercolor=green,#1]{#2}}
\colorlet{darkblue}{blue!50!black}
\hypersetup{
	colorlinks,%
	citecolor=darkblue,%
	filecolor=red,%
	linkcolor=darkblue,%
	urlcolor=darkblue,%
	pdfnewwindow=true,%
	pdfstartview={FitH}
}

\numberwithin{equation}{section}

\def\init{\setcounter{equation}{0}}

\newcounter{smallarabics}

\newcounter{smallroman}

\newcommand{\ben}{\begin{enumerate}[{\rm (1)}]}
\newcommand{\een}{\end{enumerate}}

\newtheorem{theorem}{Theorem}[section]
\newtheorem{proposition}[theorem]{Proposition}
\newtheorem{lemma}[theorem]{Lemma}

\newtheorem{corollary}[theorem]{Corollary}

\usepackage[normalem]{ulem}

\NewDocumentCommand{\ASS}{mm}{\expandafter\newcommand\csname #1\endcsname{{\hyperref[#1]{\bf (#2)}}}}
\NewDocumentCommand{\preASS}{mm}{\expandafter\newcommand\csname pre#1\endcsname{{\hyperref[#1]{\bf (#2)}}}}

\AtBeginEnvironment{theorem}{\Needspace{5\baselineskip}}
\AtBeginEnvironment{proposition}{\Needspace{3\baselineskip}}
\AtBeginEnvironment{definition}{\Needspace{5\baselineskip}}
\AtBeginEnvironment{corollary}{\Needspace{5\baselineskip}}
\AtBeginEnvironment{lemma}{\Needspace{5\baselineskip}}
\AtBeginEnvironment{quote}{\Needspace{5\baselineskip}}

 \def\cH{{\mathcal H}}

\def\d{\mathrm{d}}
\def\sing{\mathrm{sing}}

\def\bep{\begin{proposition}}
\def\eep{\end{proposition}}
\def\bet{\begin{theorem}}
\def\eet{\end{theorem}}
\def\bel{\begin{lemma}}
\def\eel{\end{lemma}}

\def\Im{\mathop{\mathrm{Im}}}

\newcommand{\one}{\mathbbm{1}}

\def\proof{\noindent {\bf Proof.}\ \ }
\def\qed{\hfill $\Box$\medip}

\def\textsl{{}}

\def\c0inf{C_0^\infty}
\def\proof{\noindent {\bf Proof.}\ \ }

\newcommand{\beq}{\begin{equation}}
\newcommand{\eeq}{\end{equation}}
\newcommand{\bear}[1]{\begin{array}{#1}}
\newcommand{\ear}{\end{array}}

\renewcommand{\d}{\mathrm{d}}

\def\qed{$\Box$\medskip}

\def\supp{{\rm supp}}
\def\ac{{\rm ac}}

\def\rr{{\mathbb R}}
\def\zz{{\mathbb Z}}
\def\cc{{\mathbb C}}

\def\P{{\mathbb P}}
\def\E{{\mathbb E}}
\def\cC{{\mathcal C}}

\def\bec{\begin{corollary}}
\def\eec{\end{corollary}}

\begin{document}
\def\today{}
\title{A Note on Extended States on the Bethe Lattice}
\author{Vojkan  Jak\v{s}i\'c$^{1}$ \,\,\,Yoram  Last$^{2}$\,\,\, Simone  Warzel$^{3}$
\\ \\
$^1$Dipartimento di Matematica\\
Politecnico di Milano, 
Piazza Leonardo da Vinci 32\\
20133 Milano, Italy
\\ \\
$^2$Einstein Institute of Mathematics\\
The Hebrew University of Jerusalem\\
Edmond J. Safra Campus, Givat Ram\\
9190401 Jerusalem, Israel
\\ \\
$^3$Department of Mathematics\\
Technical University of Munich,  Boltzmannstr. 3\\
85748 Garching b. M\"unchen, Germany
}
\date{}
\maketitle
\thispagestyle{empty}

\begin{center}
\IfFileExists{Stas_1.JPG}{\includegraphics[width=0.62\textwidth]{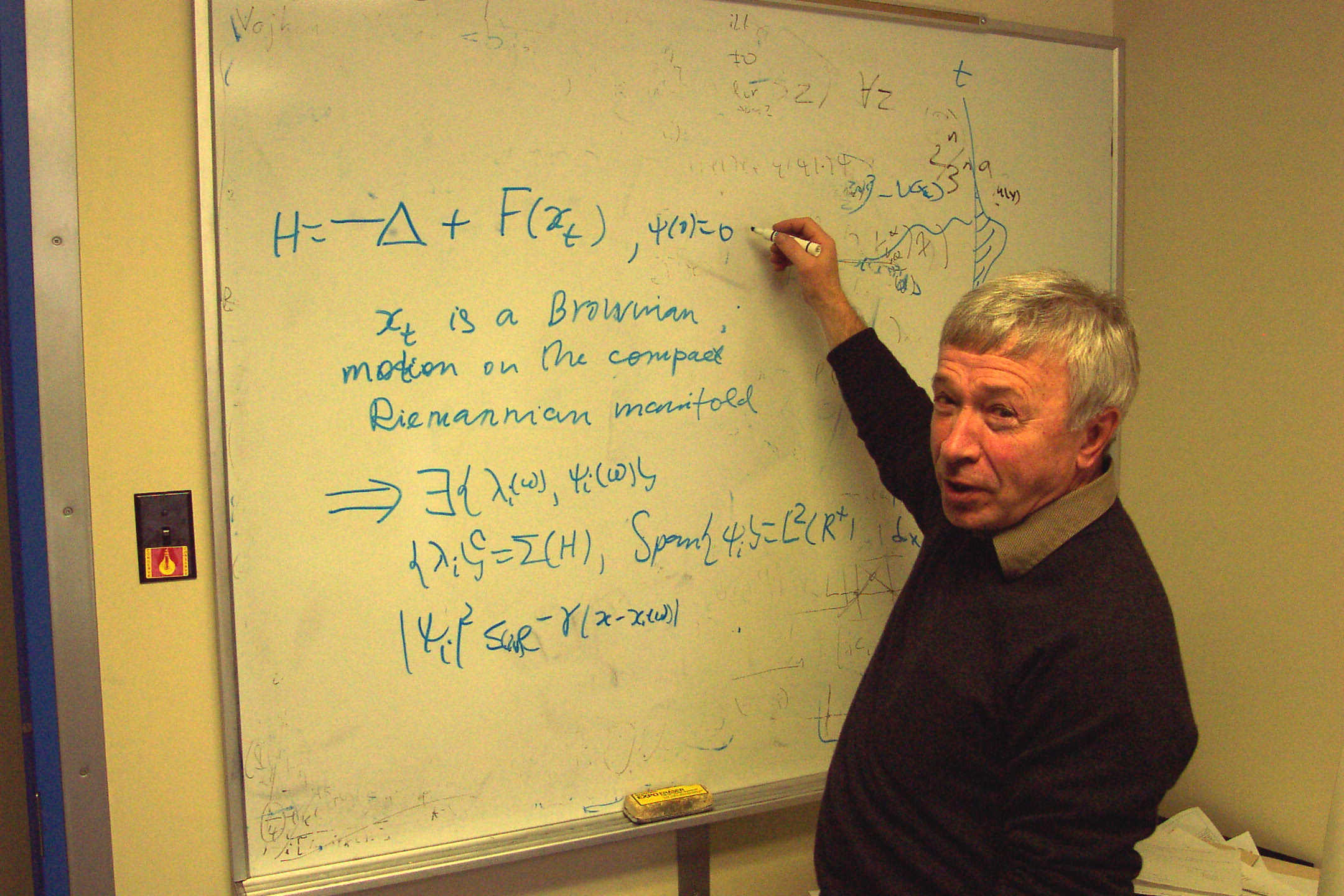}}{}


{\itshape
Dedicated to Stanislav Molchanov, on the occasion of his 85th birthday.
}

\end{center}

{\small
{\bf Abstract.} We give a proof of existence of absolutely continuous spectrum for the weakly disordered Anderson model on the Bethe lattice which uses the cyclicity criterion of \cite{JLsimple}.  The proof is based on a Hellinger overlap estimate for the spectral measures of two independent copies of the rooted tree.  The required weak disorder estimate follows from a short harmonic and compactness argument.  
We also formulate an analogous quantitative cyclicity criterion for the Anderson model on $\zz^d$.  It is expressed through the Schur complement of the Poisson transform of a $2\times2$ matrix spectral measure.
}

\vspace{1.5em}

\section{Introduction}
\init
We give a proof of existence of absolutely continuous spectrum for the
weakly disordered Anderson model on the Bethe lattice. The argument
reduces the problem to overlap between the spectral measures of independent
forward branches, which forces the root vector to be non-cyclic.

Let $\mathbb G$ be a connected graph of bounded degree, and let $A$
be its adjacency operator. Consider the Anderson type Hamiltonian
$H_\omega=A+V_\omega$ on $\ell^2(\mathbb G)$, where the
$V_\omega(x)$, $x\in\mathbb G$, are independent random variables
with absolutely continuous probability distributions. Let
$\cC_\omega(\delta_x)$ denote the cyclic subspace generated by
$H_\omega$ and $\delta_x$.
 By Theorem 1.2
and Remark 2 of \cite{JLsimple}, for every $x\in\mathbb G$,
\[
\cH_{\omega,\sing}\subset \cC_\omega(\delta_x)
\]
for $\P$-a.e.\ $\omega$.  Hence
\beq
\cC_\omega(\delta_x)^\perp\subset \cH_{\omega,\ac}.
\label{orth-ac}
\eeq
As observed in the discussion following Theorem 1.2 of
\cite{JLsimple}, it  suffices to prove that $\delta_x$ is
non-cyclic with positive probability in order to conclude that
$H_\omega$ has nontrivial absolutely continuous spectrum for
$\P$-a.e.\ $\omega$.

As in~\cite{ASW}, we work with the regular rooted tree $\mathbb T$ of
branching number $K\geq2$, with root denoted by $0$.
If the spectral measures of two
independent forward branches are not mutually singular, then the root is not a
cyclic vector. Let $\mu_\omega$ be the spectral measure of $\delta_0$ for
$H_{\lambda,\omega}=A+\lambda V_\omega$.  The  Green
function at the root is 
\[
\Gamma_\lambda(z)
=\langle\delta_0,(H_{\lambda,\omega}-z)^{-1}\delta_0\rangle
=\int_\rr\frac{\d\mu_\omega(t)}{t-z},
\qquad z\in\cc_+.
\]
For every compact interval $I\Subset(-2\sqrt K,2\sqrt K)$, we show that, for sufficiently small $|\lambda|$,
\[
\liminf_{\eta\downarrow0}
\int_I\left(\E\sqrt{\Im\Gamma_\lambda(E+\mathrm{i}\eta)}\right)^2\d E>0.
\]
By Lemma~\ref{lemma-affinity-limit}, this estimate implies that the
expectation of the Hellinger affinity of two independent root spectral measures
is positive. Hence these measures are not mutually singular with
positive probability. The root decomposition then gives non-cyclicity
of $\delta_0$, and the criterion of~\cite{JLsimple} 
yields almost sure existence of nontrivial absolutely continuous spectrum. 
As shown in Section~\ref{sec-tree-ac}, the argument also gives  nontrivial
absolutely continuous spectrum in $I$.

We also formulate a corresponding criterion for the Anderson model on
$\zz^d$ in terms of the resolvent matrix at two neighboring sites;
see Section~\ref{sec-lattice-criterion}.

For the rooted tree  and $\lambda\ne 0$ there is a useful strengthening of the criterion
recalled above.  The following three statements are equivalent:
\begin{enumerate}[(i)]
\item
$H_{\lambda,\omega}$ has nontrivial absolutely continuous spectrum for
$\P$-a.e.\ $\omega$;

\item
$\delta_0$ is non-cyclic for $H_{\lambda,\omega}$ with positive
probability;

\item
the root spectral measures of two independent copies of $\mathbb T$ are not
mutually singular with positive probability.
\end{enumerate}
The implication (ii)$\Rightarrow$(i) is the criterion stated in the
discussion following Theorem~1.2 of~\cite{JLsimple}.
By Corollary~1.1.1 of~\cite{JLstructure}, the root spectral measure
belongs almost surely to the spectral measure class of the operator.
Corollary~1.1.3 shows that the essential support of the absolutely continuous spectrum  is nonrandom.  If (i) holds, this support has positive
Lebesgue measure. Hence the root spectral measures of two independent
copies of $\mathbb T$ are not mutually singular for almost every pair
of realizations. This proves (i)$\Rightarrow$(iii).  Finally, removing the root leaves $K$ independent copies of
$\mathbb T$, so (iii)$\Rightarrow$(ii) follows from
Lemma~\ref{lemma-root-cyclic}.

By Corollary~1.1.2 of~\cite{JLstructure}, the singular parts of the
spectral measures of two independent realizations are mutually singular
almost surely. Thus the overlap in (iii) detects precisely the absolutely
continuous component. Theorem~\ref{theorem-hellinger-bound} verifies this
overlap condition at weak disorder.

The same decoupling argument shows that the absolutely continuous spectrum
on $\mathbb T$, whenever present, has infinite multiplicity.  Indeed,
decoupling the exterior of a ball of radius $n$ produces $K^{n+1}$ copies
of $\mathbb T$ with the same absolutely continuous essential  support.
By the Kato--Rosenblum theorem the multiplicity is therefore at least
$K^{n+1}$, and hence infinite.

The main result of this note---the existence of absolutely continuous
spectrum at weak disorder on the Bethe lattice---is not new.  Much
stronger results are available in the literature.  Klein
\cite{Klein94,Klein98} proved purely absolutely continuous
spectrum on compact intervals in the interior of the free spectrum.
Stability of the absolutely continuous spectral density in a more general
tree setting was established in \cite{ASW}, while the
resonant-delocalization mechanism of \cite{AW}  reaches spectral
regimes beyond the free spectrum.

The purpose of the present note is different.  We give a minimal proof of
existence which follows the necessary-and-sufficient route described
above and stops as soon as the Hellinger overlap is established. At its core, the weak-disorder estimate uses a qualitative version of the
fluctuation-suppression mechanism developed in \cite{ASW}. The present argument provides no information
about purity, regularity of the spectral density, or mobility edges.
We hope nevertheless that this minimal approach sheds some light on the
earlier results.  More speculatively, it may also be of relevance to the study  of extended states
for the Anderson model on $\zz^d$.

After the present note was submitted, the first two authors gave
in~\cite{JLrevisited} a short proof that, at sufficiently weak disorder,
the spectrum is purely absolutely continuous on compact intervals
in the interior of the free spectrum. That proof follows a completely
different strategy, based on stability of the forward Green function
and the boundary zero--one law.

Complementary localization results on the Bethe lattice were obtained by
Aizenman and Molchanov using the fractional moment method~\cite{AM};
cf.\ \cite{AWbook} for an overview.

It is a pleasure to dedicate this note to Stanislav Molchanov on the occasion of his 85th birthday. 
The first author met Stas at Caltech in the spring of 1990. In the years since, 
Stas has been a mentor, teacher, collaborator, and, above all, a close friend. 
They have shared mathematics, hiked, fished, and cooked together.
V.J. thanks Stas for it all.

\paragraph*{Acknowledgments.}
This work was partly funded by the CY Initiative grant
\emph{Investissements d'Avenir}, grant number ANR-16-IDEX-0008.
V.J. acknowledges the support of the MUR grant
``Dipartimento di Eccellenza 2023--2027'' of the Dipartimento di Matematica,
Politecnico di Milano.
We also acknowledge the support of the ANR project DYNACQUS,
grant number ANR-24-CE40-5714. This note was written while the first and third authors, V.J. and S.W., were
participating in the Les Houches Summer School \emph{Quantum Theory on All
Scales}, August 3--28, 2026. They wish to thank the organizers
S. Bachmann, S. Cenatiempo, A. Joye, and M. Salmhofer, the staff of the
Les Houches Physics School, and all the other participants for the wonderful
scientific atmosphere and for making the school such an enjoyable place. 

\section{Cyclicity and Hellinger overlap}
\label{sec-cyclicity-hellinger}
\init

Let $B_j$ be self-adjoint operators on Hilbert spaces $\cH_j$, let $e_j\in
\cH_j$, and set on $\oplus_{j}\cH_j$
\begin{equation*}
B=\bigoplus_{j=1}^N B_j,
\qquad
u=\bigoplus_{j=1}^N e_j.
\end{equation*} 
Let $a\in\rr$ and consider on $\cc\oplus\bigoplus_j\cH_j$ the operator
\begin{equation*}
H=
\begin{pmatrix}
a&\langle u,\cdot\rangle\\
u&B
\end{pmatrix}.
\end{equation*}

\bel
With the notation above,
\beq
\cC_H(1\oplus0)=\cc\oplus\cC_B(u).
\label{cyclic-root-identity}
\eeq
If the spectral measures of $(B_i,e_i)$ and $(B_j,e_j)$ are not mutually
singular for some $i\ne j$, then $1\oplus0$ is not cyclic for $H$.
\label{lemma-root-cyclic}
\eel

\proof
The space on the right-hand side of (\ref{cyclic-root-identity}) reduces
$H$ and contains $1\oplus0$. Conversely,
\[
H(1\oplus0)-a(1\oplus0)=0\oplus u,
\]
so $\cC_H(1\oplus0)$ contains both $1\oplus0$ and $0\oplus u$.
It therefore reduces the bounded off-diagonal coupling. Since it also
reduces $H$, it reduces $a\oplus B$ and contains
$0\oplus\cC_B(u)$. This proves~\eqref{cyclic-root-identity}.

Let $\mu_i$ and $\mu_j$ be the two spectral measures,  set
$\rho=\mu_i+\mu_j$, and abbreviate $w_k=\d\mu_k/\d\rho$.  If
$D=\{w_iw_j>0\}$ has positive $\rho$-measure, then
\[
0 \neq \psi_i=\boldsymbol{1}_D w_j\in L^2(\mu_i),
\qquad
0 \neq \psi_j=-\boldsymbol{1}_D w_i\in L^2(\mu_j).
\]
The vector with components $\psi_i,\psi_j$ and all other components equal to
zero is nonzero and is orthogonal to $f(B)u$ for every bounded Borel
function $f$.  Thus $u$ is not cyclic for $B$, and the assertion follows
from (\ref{cyclic-root-identity}).
\qed

For two probability measures $\mu$ and $\nu$ on $\rr$, the Hellinger affinity is
\begin{equation*}
\mathcal A(\mu,\nu)=
\int_\rr
\sqrt{\frac{\d\mu}{\d\rho}\frac{\d\nu}{\d\rho}}\,\d\rho,
\qquad
\rho=\mu+\nu.
\end{equation*}
Then
\beq
\mathcal A(\mu,\nu)>0
\quad\Longleftrightarrow\quad
\mu\not\perp\nu.
\label{affinity-overlap}
\eeq
Let
\[
P_\eta(x)=\frac1\pi\frac{\eta}{x^2+\eta^2}
\]
be the Poisson kernel.  We denote the Poisson transform of $\mu$ by $P_\eta*\mu$.  Set
\begin{equation*}
\mathcal A_\eta(\mu,\nu)
=
\int_\rr\sqrt{(P_\eta*\mu)(E)(P_\eta*\nu)(E)}\,\d E.
\end{equation*}

\bel
For any two probability measures $\mu$ and $\nu$,
\beq
\lim_{\eta\downarrow0}\mathcal A_\eta(\mu,\nu)
=\mathcal A(\mu,\nu).
\label{affinity-limit}
\eeq
\label{lemma-affinity-limit}
\eel

\proof
Applying the same Markov kernel to two measures can only increase their
Hellinger affinity, and so $\mathcal A_\eta(\mu,\nu)\geq\mathcal A(\mu,\nu)$. 
One can also argue as follows. Let $\alpha$ be the Hellinger measure,
\[
\d\alpha
=\sqrt{\frac{\d\mu}{\d\rho}\frac{\d\nu}{\d\rho}}\,\d\rho.
\]
Cauchy--Schwarz gives
\[
(P_\eta*\alpha)(E)
\leq\sqrt{(P_\eta*\mu)(E)(P_\eta*\nu)(E)}.
\]
Integrating in $E$ and using $\int P_\eta=1$ yields $\mathcal A_\eta(\mu,\nu)\geq\mathcal A(\mu,\nu)$.

On the other hand, Hellinger affinity is upper semicontinuous under weak
convergence.  One convenient way to see this is the variational formula
\[
\mathcal A(\mu,\nu)
=
\inf_{h}
\frac12\left(\int h\,\d\mu+\int h^{-1}\,\d\nu\right),
\]
where the infimum can be taken over bounded positive continuous functions
which are bounded away from zero.  Since $P_\eta*\mu\rightharpoonup\mu$ and
$P_\eta*\nu\rightharpoonup\nu$, we obtain
\[
\limsup_{\eta\downarrow0}\mathcal A_\eta(\mu,\nu)
\leq\mathcal A(\mu,\nu).
\]
This proves (\ref{affinity-limit}).
\qed

{\bf Remark.}
The proof also shows that the measures
$\sqrt{(P_\eta*\mu)(E)(P_\eta*\nu)(E)}\,\d E$
converge weakly to the Hellinger measure $\alpha$. Indeed, the difference
between their densities and $P_\eta*\alpha$ is nonnegative and its
integral tends to zero. Thus, for every compact interval $I$,
\beq
\limsup_{\eta\downarrow0}
\int_I\sqrt{(P_\eta*\mu)(E)(P_\eta*\nu)(E)}\,\d E
\leq\alpha(I).
\label{affinity-local-limit}
\eeq
We will use this estimate in Section~\ref{sec-tree-ac} to obtain a
nontrivial absolutely continuous spectrum  in $I$.

\section{The Anderson model on the Bethe lattice}
\label{sec-tree-model}
\init
\subsection{Setup and statement}
Let $K\geq2$, and let $\mathbb T$ be the regular rooted tree of branching
number $K$, with root $0$, so that every vertex has $K$ forward neighbors.
We work with this rooted version of the Bethe lattice throughout.
The corresponding result for the $(K+1)$-regular Bethe lattice follows
from the Kato--Rosenblum theorem. Indeed, decoupling one vertex leaves
$K+1$ copies of $\mathbb T$ and an isolated vertex, and changes the
Hamiltonian by a finite-rank operator.

We assume that the random variables $V_\omega(x)$ are identically
distributed and that their common distribution has compact support.

Recall that $\Gamma_\lambda(z)=\langle\delta_0,(H_{\lambda,\omega}-z)^{-1}\delta_0\rangle$. 
Let $\Gamma_{\lambda,1}(z),\ldots,\Gamma_{\lambda,K}(z)$ be independent
copies of $\Gamma_\lambda(z)$, and let $V_0$ be a copy of
$V_\omega(0)$, independent of them. Define
\beq
\Gamma_{\lambda,0}(z)
=
\frac1{\lambda V_0-z-\sum_{j=1}^K\Gamma_{\lambda,j}(z)}.
\label{root-resolvent-recursion}
\eeq
By self-similarity, $\Gamma_{\lambda,0}(z)$ has the same distribution as
$\Gamma_\lambda(z)$.
For $\lambda=0$, the root Green function is the deterministic solution
$g(z)\in\cc_+$ of
\begin{equation*}
g(z)=\frac1{-z-Kg(z)}.
\end{equation*}
In particular,
\begin{equation*}
g(E+\mathrm{i}0)
=\frac{-E+\mathrm{i}\sqrt{4K-E^2}}{2K},
\qquad |E|<2\sqrt K,
\end{equation*}
and
\beq
\sqrt K\,|g(E+\mathrm{i}0)|=1.
\label{free-modulus}
\eeq

\bet
Let $I\Subset(-2\sqrt K,2\sqrt K)$ be a compact interval of positive length. There are
$c_I>0$ and $\lambda_I>0$ such that, for every $|\lambda|<\lambda_I$,
\beq
\liminf_{\eta\downarrow0}
\int_I
\left(\E\sqrt{\Im\Gamma_\lambda(E+\mathrm{i}\eta)}\right)^2\d E
\geq c_I.
\label{main-hellinger-lower-bound}
\eeq
\label{theorem-hellinger-bound}
\eet

\bec
Let $I$ and $\lambda_I$ be as in Theorem~\ref{theorem-hellinger-bound}.
For every $|\lambda|<\lambda_I$, the Anderson model
$H_{\lambda,\omega}$ on $\mathbb T$ has a nontrivial absolutely continuous
spectrum  in $I$ for $\P$-a.e.\ $\omega$.
\label{corollary-tree-ac}
\eec

The proof of Theorem~\ref{theorem-hellinger-bound} occupies the next two
subsections.

\subsection{The Lyapunov function}

Set
\[
Y_{\lambda,j}(z)=\Im\Gamma_{\lambda,j}(z),
\qquad j=0,1,\ldots,K.
\]
Taking imaginary parts in (\ref{root-resolvent-recursion}) gives
\beq
Y_{\lambda,0}(z)
=
|\Gamma_{\lambda,0}(z)|^2
\left(
\Im z+\sum_{j=1}^K Y_{\lambda,j}(z)
\right).
\label{imaginary-recursion}
\eeq
The random variables $Y_{\lambda,1},\ldots,Y_{\lambda,K}$ are
independent copies of $Y_\lambda=\Im\Gamma_\lambda$.
The variable $Y_{\lambda,0}$ has the same distribution.  
Set
\[
\gamma_\lambda(z)
=
-\E\log\big(\sqrt K|\Gamma_\lambda(z)|\big).
\]
Since $\Gamma_{\lambda,0}$ has the same distribution as $\Gamma_\lambda$,
taking logarithms in (\ref{imaginary-recursion}) and using equality of the
marginal laws gives
\beq
2\gamma_\lambda(z)
=
\E\left[
\log\frac{\Im z+\sum_{j=1}^K Y_{\lambda,j}(z)}{K}
-\frac1K\sum_{j=1}^K\log Y_{\lambda,j}(z)
\right].
\label{jensen-gap-identity}
\eeq
The arithmetic-geometric mean inequality gives $\gamma_\lambda(z)\geq0$.

The following elementary observation is the only concentration argument
which we shall need.

\bel
Let $\eta_n\geq0$ and let $Y_n$ be positive random variables.  Let
$Y_{n,1},\ldots,Y_{n,K}$ be independent copies of $Y_n$ and suppose that
\beq
\E\left[
\log\frac{\eta_n+\sum_{j=1}^K Y_{n,j}}
{K\left(\prod_{j=1}^K Y_{n,j}\right)^{1/K}}
\right]\rightarrow0.
\label{one-scale-assumption}
\eeq
Then there are numbers $a_n>0$ such that
\beq
\frac{Y_n}{a_n}\rightarrow1
\quad\hbox{in probability and }
\quad
\frac{\eta_n}{a_n}\rightarrow0.
\label{one-scale-conclusion}
\eeq
\label{lemma-one-scale}
\eel

\proof
Set
\[
S_n=\sum_{j=1}^K Y_{n,j},
\qquad
q_{n,j}=\frac{Y_{n,j}}{S_n}.
\]
The random variable in (\ref{one-scale-assumption}) is nonnegative and can
be written as
\begin{equation*}
\log\left(1+\frac{\eta_n}{S_n}\right)
-\frac1K\log\left(K^K\prod_{j=1}^Kq_{n,j}\right) \geq 0 .
\end{equation*}
By the arithmetic-geometric mean inequality the second term is also non-negative. 
Since the expectation of their
sum tends to zero, Markov's inequality implies 
\beq
\frac{\eta_n}{S_n}\rightarrow0,
\qquad
K^K\prod_{j=1}^Kq_{n,j}\rightarrow1
\quad\hbox{in probability}.
\label{gap-consequences}
\eeq
On the simplex $\sum_jq_j=1$, the product $\prod_jq_j$ has the unique
maximum $K^{-K}$ at $q_1=\cdots=q_K=1/K$.  Thus, for every
$\varepsilon>0$, there is $\delta_\varepsilon>0$ such that
\[
\max_j\left|q_j-\frac1K\right|\geq\varepsilon
\quad\Rightarrow\quad
K^K\prod_{j=1}^Kq_j\leq1-\delta_\varepsilon.
\]
The second relation in (\ref{gap-consequences}) therefore gives
\begin{equation*}
q_{n,j}\rightarrow\frac1K,
\qquad j=1,\ldots,K,
\quad\hbox{in probability}.
\end{equation*}
In particular,
\[
\frac{Y_{n,1}}{Y_{n,2}}
=
\frac{q_{n,1}}{q_{n,2}}
\rightarrow1
\quad\hbox{in probability}.
\]

Let $c_n$ be a median of $\log Y_n$. Independence and the median
property give, for every $\varepsilon>0$,
\[
\P\{|\log Y_n-c_n|>\varepsilon\}
\leq
2\P\{|\log Y_{n,1}-\log Y_{n,2}|>\varepsilon\}
\rightarrow0.
\]
Thus $\log Y_n-c_n\rightarrow0$ in probability. With $a_n=e^{c_n}$, it follows that
\[
\frac{Y_n}{a_n}\rightarrow1
\quad\hbox{in probability}.
\]
Consequently $S_n/a_n\to K$ in probability, and the first relation in
(\ref{gap-consequences}) yields
\[
\frac{\eta_n}{a_n}
=
\frac{\eta_n}{S_n}\frac{S_n}{a_n}
\rightarrow0.
\]
This proves (\ref{one-scale-conclusion}).
\qed

We also need the following elementary lemma.

\bel
Let $u_n,u$ be nonnegative harmonic functions on $\cc_+$ such that
$u_n\to u$ locally uniformly.  Suppose that $u$ extends continuously to an
open interval $J\subset\rr$ and vanishes there.  Then, for every compact
$I\Subset J$ and every sequence $\eta_n\downarrow0$,
\beq
\int_Iu_n(E+\mathrm{i}\eta_n)\d E\rightarrow0.
\label{harmonic-boundary-limit}
\eeq
\label{lemma-harmonic-boundary}
\eel

\proof
Choose $I\Subset J_0\Subset J$ and fix $y>0$ sufficiently small.  For all
large $n$, $\eta_n<y$.  Positivity and the Poisson formula in the half-plane
above the line $\Im z=\eta_n$ give
\[
u_n(x+\mathrm{i}y)
\geq
\int_\rr P_{y-\eta_n}(x-t)u_n(t+\mathrm{i}\eta_n)\d t.
\]
Since $I\Subset J_0$, for all sufficiently small $y$ and all large $n$,
\[
\inf_{t\in I}\int_{J_0}P_{y-\eta_n}(x-t)\d x\geq\frac12.
\]
Hence
\[
\int_Iu_n(t+\mathrm{i}\eta_n)\d t
\leq
2\int_{J_0}u_n(x+\mathrm{i}y)\d x.
\]
Taking first $n\to\infty$ and then $y\downarrow0$ proves
(\ref{harmonic-boundary-limit}).
\qed

\bel
For every compact interval $I\Subset(-2\sqrt K,2\sqrt K)$,
\begin{equation*}
\lim_{\substack{\lambda\to0\\ \eta\downarrow0}}
\int_I\gamma_\lambda(E+\mathrm{i}\eta)\d E=0.
\end{equation*}
\label{lemma-integrated-gamma}
\eel
\proof
Since the single-site potential is bounded, $\Gamma_\lambda(z)\rightarrow g(z)$ 
locally uniformly on $\cc_+$, uniformly in the disorder realization.
Consequently,
\[
\gamma_\lambda(z)\rightarrow
\gamma_0(z)
=
-\log(\sqrt K|g(z)|)
\]
locally uniformly on $\cc_+$.  The functions $\gamma_\lambda$ are
nonnegative harmonic and, by (\ref{free-modulus}), $\gamma_0$ extends
continuously to zero on $(-2\sqrt K,2\sqrt K)$.  Given arbitrary sequences
$\lambda_n\to0$ and $\eta_n\downarrow0$, the assertion follows from
Lemma \ref{lemma-harmonic-boundary} with
$u_n=\gamma_{\lambda_n}$ and $u=\gamma_0$.
\qed

\subsection{Proof of Theorem \ref{theorem-hellinger-bound}}

Suppose that the assertion is false.  Then there are sequences $\lambda_n\rightarrow0$, $\eta_n\downarrow0$, 
such that
\begin{equation*}
\int_I
\left(
\E\sqrt{\Im\Gamma_{\lambda_n}(E+\mathrm{i}\eta_n)}
\right)^2\d E
\rightarrow0.
\end{equation*}
Together with Lemma \ref{lemma-integrated-gamma}, this allows us to choose
$E_n\in I$ such that, with
\[
z_n=E_n+\mathrm{i}\eta_n,
\qquad
X_n=\Gamma_{\lambda_n}(z_n),
\qquad
Y_n=\Im X_n,
\]
one has
\beq
\E\sqrt{Y_n}\rightarrow0,
\qquad
\gamma_{\lambda_n}(z_n)\rightarrow0.
\label{two-small-quantities}
\eeq
Passing to a subsequence, we may assume that $E_n\to E\in I$.

By (\ref{jensen-gap-identity}) and Lemma \ref{lemma-one-scale}, there is 
a sequence  $a_n>0$ such that
\beq
\frac{Y_n}{a_n}\rightarrow1
\quad\hbox{in probability and}
\quad
\frac{\eta_n}{a_n}\rightarrow0.
\label{hellinger-one-scale}
\eeq
Since $\E\sqrt{Y_n}\rightarrow0$ by (\ref{two-small-quantities}) and
$\P\{Y_n\geq a_n/2\}\rightarrow1$ by (\ref{hellinger-one-scale}),
we have $a_n\rightarrow0$.

Let $X_{n,1},\ldots,X_{n,K}$ be independent copies of $X_n$, and let
$V_{n,0}$ be a copy of $V_0$, independent of them.  Set
\[
X_{n,0}
=
\frac1{\lambda_nV_{n,0}-z_n-\sum_{j=1}^KX_{n,j}},
\qquad
Y_{n,j}=\Im X_{n,j},\quad j=0,1,\ldots,K.
\]
By self-similarity of the rooted tree and the root resolvent recursion,
$X_{n,0}$ has the same law as $X_n$. Hence
\[
\frac{Y_{n,j}}{a_n}\rightarrow1,
\qquad j=0,1,\ldots,K,
\]
in probability.  The relation~\eqref{imaginary-recursion} yields
\[
|X_{n,0}|^2
=
\frac{Y_{n,0}/a_n}
{\eta_n/a_n+\sum_{j=1}^K Y_{n,j}/a_n},
\]
and so $|X_{n,0}|^2\to1/K$ in probability.  Since $X_{n,0}$ has the
same law as $X_n$, $|X_n|^2\rightarrow\frac1K$ in probability.
Since $a_n\rightarrow0$ and $Y_n/a_n\rightarrow1$ in probability,
we also have $Y_n\rightarrow0$ in probability. Thus
\begin{equation*}
{\rm dist}\left(
X_n,\left\{-\frac1{\sqrt K},\frac1{\sqrt K}\right\}
\right)
\rightarrow0
\quad\hbox{in probability}.
\end{equation*}

Passing to a subsequence, the laws of $X_n$ converge to
\beq
\nu
=
p\delta_{1/\sqrt K}
+
(1-p)\delta_{-1/\sqrt K}
\label{two-point-law}
\eeq
for some $p\in[0,1]$.  By the definition of $X_{n,0}$,
\beq
\left(
\lambda_nV_{n,0}-E_n-\mathrm{i}\eta_n
-\sum_{j=1}^KX_{n,j}
\right)X_{n,0}=1.
\label{polynomial-recursion-n}
\eeq
Passing to a further subsequence if necessary, the joint laws of
$(X_{n,0},X_{n,1},\ldots,X_{n,K})$ converge weakly.  Let
$(X_0,X_1,\ldots,X_K)$ have the limiting joint law.  From
(\ref{polynomial-recursion-n}),
\[
\left(
-E_n-\mathrm{i}\eta_n-\sum_{j=1}^KX_{n,j}
\right)X_{n,0}-1
=
-\lambda_nV_{n,0}X_{n,0}.
\]
The single-site potential is bounded, $\lambda_n\to0$, and $(X_{n,0})$
is tight.  Hence the right-hand side tends to zero in probability.
Since also $E_n\to E$ and $\eta_n\to0$, it follows that
\beq
\left(
-E-\sum_{j=1}^KX_j
\right)X_0=1
\quad\hbox{almost surely}.
\label{limiting-recursion}
\eeq
The variables $X_1,\ldots,X_K$ are independent with common law $\nu$. 
$X_0$ has marginal law $\nu$ but is not assumed to be independent
of $(X_1,\ldots,X_K)$. If $0<p<1$, then $X_1+\cdots+X_K$ takes each of the $K+1$ distinct values $(2m-K)/\sqrt K$, $m=0,\ldots,K$, 
with positive probability.  Since the map $s\mapsto(-E-s)^{-1}$ 
is injective, (\ref{limiting-recursion}) would force the law of $X_0$ to
have at least $K+1\geq3$ points in its support.  This contradicts
(\ref{two-point-law}). Hence $\nu=\delta_x$ with $x=\pm K^{-1/2}$.  Equation
(\ref{limiting-recursion}) gives $Kx^2+Ex+1=0$. 
Since $Kx^2=1$, this implies $E=\pm2\sqrt K$, contrary to $E\in I\Subset(-2\sqrt K,2\sqrt K)$. 
This proves (\ref{main-hellinger-lower-bound}).

\subsection{Proof of Corollary \ref{corollary-tree-ac}}
\label{sec-tree-ac}

We may assume $0<|\lambda|<\lambda_I$.
The Poisson transform of the spectral measure $\mu_\omega$ is 
\[
(P_\eta*\mu_\omega)(E)
=\frac1\pi\Im\Gamma_{\lambda,\omega}(E+\mathrm{i}\eta).
\]
For two independent environments $\omega,\omega'$, let
$\alpha_{\omega,\omega'}$ be the Hellinger measure of $\mu_\omega$ and
$\mu_{\omega'}$. Since $\mathcal A_\eta(\mu_\omega,\mu_{\omega'})\leq1$,
Fubini's theorem,~\eqref{affinity-local-limit}, and the reverse Fatou
lemma give
\beq
\E_{\omega,\omega'}\alpha_{\omega,\omega'}(I)
\geq
\limsup_{\eta\downarrow0}\frac1\pi
\int_I\left(\E\sqrt{\Im\Gamma_\lambda(E+\mathrm{i}\eta)}\right)^2\d E
\geq\frac{c_I}{\pi}>0,
\label{positive-affinity-limit}
\eeq
where the last inequality is Theorem~\ref{theorem-hellinger-bound}.

Remove the root from $\mathbb T$, and let $\mu_1,\ldots,\mu_K$ be
the root spectral measures of the resulting subtrees. These measures
are independent and have the same distribution as $\mu_\omega$.
Thus, with positive probability, $\mu_1$ and $\mu_2$ have nonzero
Hellinger overlap in $I$.

On this event, apply the construction in the proof of
Lemma~\ref{lemma-root-cyclic} with $D=I\cap\{w_1w_2>0\}$.
The resulting nonzero vector $\psi$ belongs to $\cC_B(u)^\perp$ and
satisfies $\one_I(B)\psi=\psi$. Since $\cC_B(u)^\perp$ reduces $B$ and is orthogonal to $u$,
the spectral measure of $0\oplus\psi$ for $H_{\lambda,\omega}$
coincides with that of $\psi$ for $B$. 
Hence~\eqref{cyclic-root-identity} gives
\[
0\ne0\oplus\psi\in
\one_I(H_{\lambda,\omega})\cC_\omega(\delta_0)^\perp.
\]
 By~\eqref{orth-ac}, $H_{\lambda,\omega}$ has nontrivial absolutely
continuous spectrum in $I$ with positive probability. 
Nonrandomness of the essential support of the absolutely continuous spectrum gives the almost-sure conclusion.

\subsection{Unbounded single-site distributions}

Theorem~\ref{theorem-hellinger-bound} and
Corollary~\ref{corollary-tree-ac} remain valid for absolutely continuous
single-site distributions satisfying
\[
\E\log(1+|V_\omega(0)|)<\infty.
\]
This is the logarithmic moment assumption in~\cite{ASW}. 
To see this, let $|\lambda|\leq1$ and $z$ belong to a compact set
$L\Subset\cc_+$. The resolvent bound
$|\Gamma_{\lambda,j}(z)|\leq(\Im z)^{-1}$
and~\eqref{root-resolvent-recursion} give
\beq
\bigl|\log|\Gamma_\lambda(z)|\bigr|
+\bigl|\log\Im\Gamma_\lambda(z)\bigr|
\leq C_L+3\log(1+|V_\omega(0)|).
\label{b-mar}
\eeq
Thus the logarithms in~\eqref{jensen-gap-identity} are integrable, and
$\gamma_\lambda$ is nonnegative and harmonic. For every realization, $H_{\lambda,\omega}\to A$ in the strong
resolvent sense as $\lambda\to0$, and hence
$\Gamma_\lambda\to g$ locally uniformly on $\cc_+$.
By~\eqref{b-mar} and dominated convergence,
$\gamma_\lambda\to\gamma_0$ locally uniformly.
Thus Lemma~\ref{lemma-integrated-gamma} remains valid.

In the final compactness argument, $\lambda_nV_{n,0}\to0$ in probability.
Since $(X_{n,0})$ is tight, $\lambda_nV_{n,0}X_{n,0}\to0$ in probability
as well, which justifies the passage to~\eqref{limiting-recursion}.
The remaining steps use no boundedness assumption.

\section{A quantitative criterion on \texorpdfstring{$\zz^d$}{Z-d}}
\label{sec-lattice-criterion}
\init

We now consider the Anderson model $H_{\omega}=A_d+V_\omega$
on $\ell^2(\zz^d)$, where $A_d$ is the unnormalized nearest-neighbor
adjacency operator and the random variables $V_\omega(x)$, $x\in\zz^d$,
are independent and identically distributed with an absolutely
continuous distribution. Let $e$ be a nearest neighbor of $0$, and let
$P_{\omega,0}$ denote the orthogonal projection onto
$\cC_\omega(\delta_0)$.

For a nonnegative function $\chi\in C_c(\rr)$, define the cyclicity deficit
\begin{equation*}
\mathfrak D_{\omega}(\chi)
=
\left\|(1-P_{\omega,0})\chi(H_{\omega})^{1/2}\delta_e\right\|^2.
\end{equation*}
Since $\cC_\omega(\delta_0)$ is reducing for $H_{\omega}$,
\beq
\mathfrak D_{\omega}(\chi)>0
\quad\Longrightarrow\quad
\delta_0\text{ is not cyclic and }
\chi(H_{\omega})\cH_{\omega,\ac}\ne\{0\}.
\label{deficit-implies-ac}
\eeq
Indeed, if $\psi=(1-P_{\omega,0})\delta_e$, then
$\mathfrak D_\omega(\chi)=\langle\psi,\chi(H_{\omega})\psi\rangle$.
Thus $\mathfrak D_\omega(\chi)>0$ implies $\psi\neq0$, so that
$\delta_0$ is non-cyclic; moreover, by (\ref{orth-ac}),
$\psi\in\cH_{\omega,\ac}$, and hence
$\chi(H_{\omega})\cH_{\omega,\ac}\neq\{0\}$.
\subsection{The Poisson--Schur formula}

Write
\[
G_\omega(x,y;z)
=\langle\delta_x,(H_{\omega}-z)^{-1}\delta_y\rangle,
\]
and set 
\begin{equation*}
M_\omega(z)=
\begin{pmatrix}
G_\omega(0,0;z)&G_\omega(0,e;z)\\
G_\omega(e,0;z)&G_\omega(e,e;z)
\end{pmatrix},
\qquad
W_{\omega,\eta}(E)=\frac1\pi\Im M_\omega(E+\mathrm{i}\eta).
\end{equation*}
The rows and columns are indexed by $0,e$,  $W_{xy}$ denotes the
corresponding matrix entry, and  $\Im M=(M-M^*)/(2\mathrm{i})$.
For a positive $2\times2$ matrix
$W=\left(\begin{smallmatrix}a&b\\ \overline b&c\end{smallmatrix}\right)$,
define
\begin{equation*}
q(W)=
\begin{cases}
c-|b|^2/a,&a>0,\\
c,&a=0.
\end{cases}
\end{equation*}
For $W=W_{\omega,\eta}(E)$ and $\eta>0$, we have $a>0$.   Write $H=H_\omega$. Since
\[
\Im(H-E-\mathrm{i}\eta)^{-1}
=\eta(H-E+\mathrm{i}\eta)^{-1}(H-E-\mathrm{i}\eta)^{-1},
\]
$W_{\omega,\eta}(E)$ is the Gram matrix of the two vectors
\[
\sqrt{\frac\eta\pi}(H-E-\mathrm{i}\eta)^{-1}\delta_0,
\qquad
\sqrt{\frac\eta\pi}(H-E-\mathrm{i}\eta)^{-1}\delta_e.
\]
Consequently,
\begin{equation*}
q(W_{\omega,\eta}(E))
=\frac\eta\pi\min_{\zeta\in\cc}
\left\|(H-E-\mathrm{i}\eta)^{-1}\delta_e
- \zeta (H-E-\mathrm{i}\eta)^{-1}\delta_0\right\|^2.
\end{equation*}

\bel
For every nonnegative $\chi\in C_c(\rr)$,
\beq
\mathfrak D_{\omega}(\chi)
=
\lim_{\eta\downarrow0}
\int_\rr\chi(E)q(W_{\omega,\eta}(E))\d E.
\label{poisson-schur-limit}
\eeq
\label{lemma-poisson-schur}
\eel

\proof
Let
\[
\mu(B)=\langle\delta_0,E_H(B)\delta_0\rangle,
\quad
\nu(B)=\langle\delta_0,E_H(B)\delta_e\rangle,
\quad
\tau(B)=\langle\delta_e,E_H(B)\delta_e\rangle.
\]
In the spectral representation of $H$ on $\cC_\omega(\delta_0)$,
let $f\in L^2(\mu)$ represent the vector $P_{\omega,0}\delta_e$. Then 
\beq
\nu=f\mu,
\qquad
\tau=|f|^2\mu+\sigma,
\label{matrix-measure-decomposition}
\eeq
where $\sigma$ is the spectral measure of
$(1-P_{\omega,0})\delta_e$.  In particular,
\beq
\mathfrak D_{\omega}(\chi)=\int\chi\,\d\sigma.
\label{deficit-sigma}
\eeq
From
(\ref{matrix-measure-decomposition}),
\begin{equation*}
q(W_{\omega,\eta}(E))
=(P_\eta*\sigma)(E)+V_\eta(E),
\end{equation*}
where
\begin{equation*}
V_\eta
=P_\eta*(|f|^2\mu)
-\frac{|P_\eta*(f\mu)|^2}{P_\eta*\mu}
\geq 0 .
\end{equation*}
We claim that
\beq
\lim_{\eta\downarrow0}\int\chi(E)V_\eta(E)\d E=0.
\label{variance-vanishes}
\eeq
We prove this by approximating $f$ in $L^2(\mu)$ by continuous functions.
Given $\varepsilon>0$, choose $g\in C_c(\rr)$ such that
$\|f-g\|_{L^2(\mu)}^2<\varepsilon$. 
For $\eta>0$, let $\pi_{E,\eta}$ be the probability measure defined by
\[
\d\pi_{E,\eta}(t)
=
\frac{P_\eta(E-t)}{(P_\eta*\mu)(E)}\,\d\mu(t).
\]
Then
\[
V_\eta=(P_\eta*\mu)\,\mathrm{Var}_{\pi_{E,\eta}}(f).
\]
The elementary inequality
\[
\mathrm{Var}(f)
\leq \int|f-g(E)|^2\d\pi
\leq
2\int|f-g|^2\d\pi
+
2\int|g(t)-g(E)|^2\d\pi
\]
gives
\[
\int\chi V_\eta
\leq
2\|\chi\|_\infty\varepsilon
+
2\|\chi\|_\infty
\iint P_\eta(E-t)|g(t)-g(E)|^2\d E\,\d\mu(t).
\]
The last double integral tends to zero as $\eta\downarrow0$.  Indeed, for
fixed $\delta>0$, its contribution from $|E-t|<\delta$ is bounded by the
squared modulus of continuity of $g$, while the contribution from
$|E-t|\geq\delta$ tends to zero with $\eta$.  Letting first
$\eta\downarrow0$, then $\delta\downarrow0$, and finally
$\varepsilon\downarrow0$ proves (\ref{variance-vanishes}).

Since $P_\eta$ is an approximate identity,
\[
\int\chi(E)(P_\eta*\sigma)(E)\d E
\rightarrow
\int\chi\,\d\sigma.
\]
Together with (\ref{deficit-sigma}), this proves
(\ref{poisson-schur-limit}).
\qed
\Needspace{5\baselineskip}
\subsection{The criterion}
\bet
Let $\chi\in C_c(\rr)$ be nonnegative.  Suppose that
\beq
\liminf_{\eta\downarrow0}
\,\E\int_\rr\chi(E)q(W_{\omega,\eta}(E))\d E>0.
\label{lattice-quantitative-criterion}
\eeq
Then $H_\omega$ has nontrivial absolutely continuous spectrum in
$\supp\chi$ for $\P$-a.e.\ $\omega$.
\label{theorem-lattice-criterion}
\eet

\proof
Since $q(W)\leq W_{ee}$,
\[
0\leq\int\chi q(W_{\omega,\eta})
\leq\|\chi\|_\infty.
\]
By Lemma \ref{lemma-poisson-schur} and dominated convergence the left-hand side of (\ref{lattice-quantitative-criterion}) equals
$\E\mathfrak D_\omega(\chi)$.  Hence
$\mathfrak D_\omega(\chi)>0$ with positive probability.  By
(\ref{deficit-implies-ac}), this gives a nonzero absolutely continuous vector
with spectral support in $\supp\chi$.  Ergodicity of the Anderson model gives
the almost sure statement.
\qed

\bigskip 

\begin{center}
\begin{minipage}{\textwidth}
\centering
\IfFileExists{Stas_2.JPG}{\includegraphics[width=0.70\textwidth]{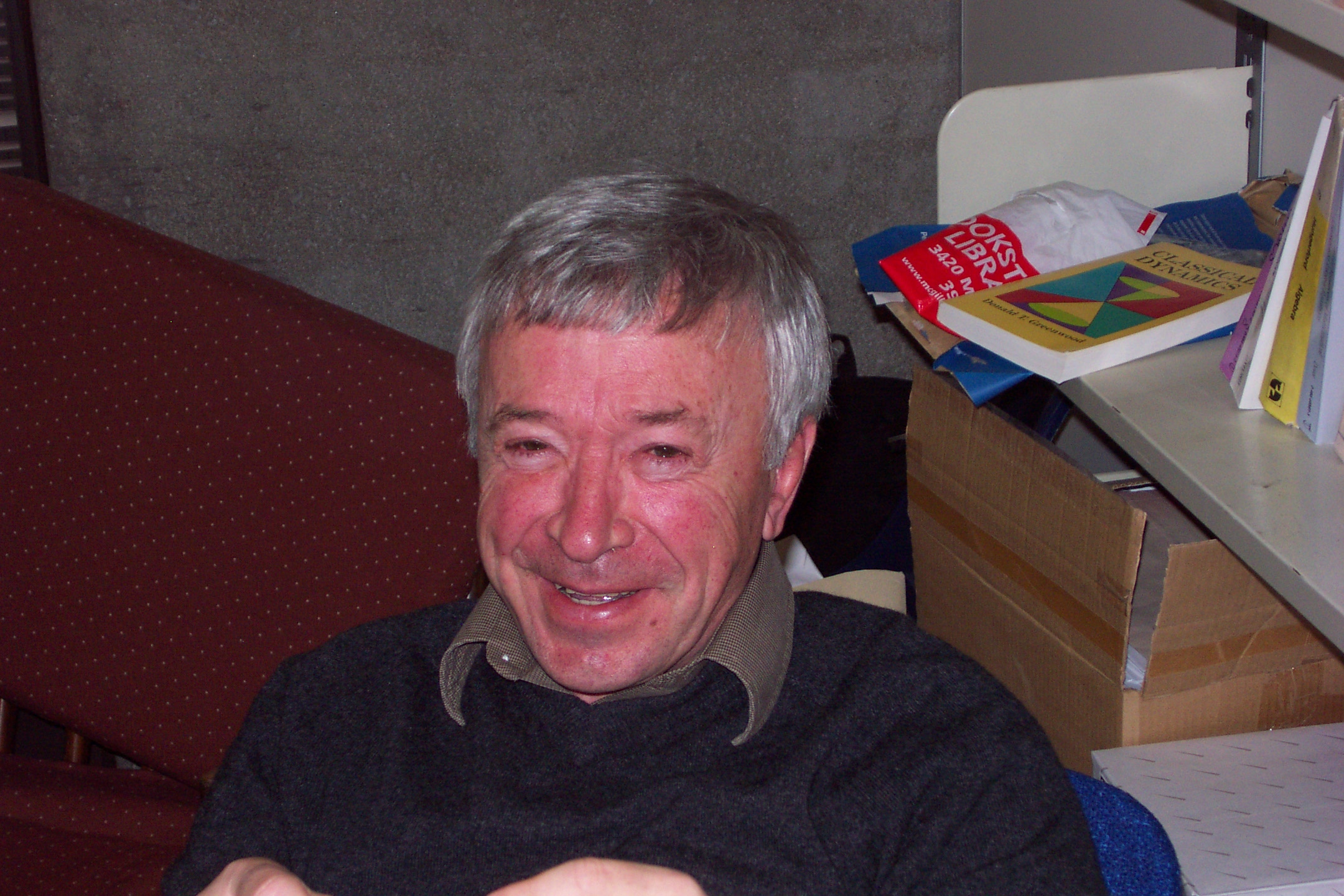}}{}

\vspace{0.8em}

{\itshape
Happy Birthday, Stas!}
\end{minipage}
\end{center}

\end{document}